\documentclass[aip,jcp,reprint,twocolumn,amsmath,amssymb,amsfonts]{revtex4-1}
\usepackage[utf8]{inputenc}
\usepackage{mathptmx}
\usepackage[scaled=.92]{helvet}

\usepackage{graphicx}
\usepackage{braket}
\usepackage{color}

\newcommand{\figref}[1]{Fig.\,\ref{#1}}
\newcommand{\tabref}[1]{Tab.\,\ref{#1}}
\newcommand{\eqeqref}[1]{Eq.\,\eqref{#1}}

\renewcommand{\vec}[1]{\mathbf{#1}}

\renewcommand{\phi}{\varphi}
\renewcommand{\epsilon}{\varepsilon}
\newcommand{\norm}[1]{\left\Vert #1 \right\Vert}

\definecolor{cdcolor}{rgb}{.6,.6,.6}

\begin{document}

\title{{\color{blue}Intrinsic atomic orbitals: An unbiased bridge between quantum theory and chemical concepts\vspace*{0.18cm}}}
\author{Gerald Knizia}
\affiliation{Institut für Theoretische Chemie, Universität Stuttgart, Pfaffenwaldring 55, D-70569 Stuttgart}

\begin{abstract}
   Modern quantum chemistry can make quantitative predictions on an immense
   array of chemical systems. However, the interpretation of those predictions is
   often complicated by the complex wave function expansions used. Here we show that
   an exceptionally simple algebraic construction allows for defining atomic core
   and valence orbitals, polarized by the molecular environment, which can exactly
   represent self-consistent field wave functions.
   This construction provides an unbiased and direct connection between quantum chemistry
   and empirical chemical concepts, and can be used, for example, to \emph{calculate}
   the nature of bonding in molecules, in chemical terms, from first principles.
\end{abstract}

\maketitle

\section{Introduction}

\newlength{\figurerefwidth}
\setlength{\figurerefwidth}{\columnwidth}

Chemical concepts as fundamental as atomic orbitals (AOs)
in molecules, covalent bonds, or even partial charges,
do not correspond to physical observables and thus cannot be unambigously defined in pure quantum theory.
This leads to the unpleasing situation that
quantum chemistry can tell us benzene's heat of formation with
<2 kJ/mol accuracy,\cite{harding:benzene}
but, strictly speaking, neither that it has twelve
localized $\sigma$-bonds and a delocalized $\pi$-system, nor what the partial
charges on the carbons are.
Chemical bonds have even
been compared to unicorns---mythical creatures of which everyone knows
how they look, despite nobody ever having seen one.\cite{frenking2007unicorns}

However, qualitative concepts are of essential importance for practical chemistry,
and thus a large number of competing techniques were developed for extracting them from quantum chemical calculations.
In particular, Bader's atoms in molecules\cite{bader:aim} and Weinholds
natural atomic/bond orbital analysis (NAO/NBO)\cite{weinhold:npa,weinhold:nho} are widely
used for interpreting molecular electronic structure.
Nonetheless, the former is known to produce counter-intuitive results in
many cases,\cite{Guerra:VDD}
and the latter, while undoubtedly having brought countless successes
in chemical interpretation, is complicated and pre-imposes
various non-trivial assumptions.
In particular, NBO analysis is based
on the two notions that atomic orbitals (AOs) in molecules have spherical symmetry
and can be
obtained by a particular complex series of transformations,\cite{weinhold:npa} and
that a Lewis-like bonding pattern for any given
molecule \emph{exists} and only needs to be found---by comparing the
wave function to all possible Lewis patterns.\cite{weinhold:nlmo}
While normally applicable,
violations of both assumptions are conceivable in unusual bonding situations,
and might then lead to erratic interpretations.

We here present a new technique to connect quantitative
self-consistent field (SCF) wave functions to a qualitative chemical picture.
This technique is essentially free of empirical input, allows for \emph{computing}
the nature and shape of chemical bonds, and is not biased towards any preconceived
notion of bonding.
This is achieved by first defining a new intrinsic minimal basis (IMB), 
a set of perturbed core- and valence AOs which can exactly describe the occupied molecular orbitals (MOs)
of a previously computed SCF wave function. We will show that the intrinsic AOs
(IAOs) thus defined can be directly interpreted as the \emph{chemical} AOs, and
that partial charges and bond orbitals (IBOs) derived from them
perfectly agree with both experimental data and intuitive chemical concepts.
In particular, we find a natural \emph{emergence} of the Lewis structure of molecules.

While other IMB constructions have been developed before,
\cite{Mayer:TheEffectiveMinimalBasis,lee:PolarizedAtomicOrbitals1,HeadGordon:PolarizedAtomicOrbitals2,auer:EnvelopingLocalizedOrbitals,Ruedenberg:MoleculeIntrinsicMinimalBasis,subotnik:LocalizedVirtuals,ruedenberg:ExactRepresentationOfRdm,laikov:IntrinsicBasis}
most are rather complex and so far none has found widespread use comparable to Bader or NAO analysis.
Our contribution is a IMB which is
simple and efficient,
its use in constructing bond orbitals,
and the demonstration that this combination shows excellent promise for
interpreting chemical bonding and reactivity.
The technique thereby provides a firm quantum mechanical
basis for ubiquitous fundamental concepts.

\section{Construction of intrinsic orbitals}
Assume that we have computed a molecular SCF wave function $\ket{\Phi}$.
$\ket{\Phi}$ is defined by its occupied MOs $\ket{i}=\sum_\mu \ket{\mu}C^\mu_i$,
where $\mu\in B_1$ are basis functions from a large basis set
$B_1$. The key problem in interpreting wave functions is that
the basis functions $\ket{\mu}$ cannot be clearly
associated with any atom; each function will
contribute most where it is most needed, and due to $B_1$'s
high variational freedom, this often is not on the atom it is placed on.
On the other hand, if one were to expand the MOs over a
minimal basis $B_2$ of free-atom AOs (i.e., a basis of accurate AOs, but e.g., with only AOs
1s,2s,2px-2pz for each C atom), the wave function would be easy to
interpret. 
But it would be inaccurate, and might be even qualitatively incorrect, because
free-atom AOs contain no polarization due to the molecular environment.
We thus propose to first calculate an accurate wave function $\ket{\Phi}$,
and then to form a set of polarized AOs $\ket{\rho}\notin B_2$
which can exactly express $\ket{\Phi}$s occupied MOs $\ket{i}$.
For this, we first split the free-atom AOs $\ket{\tilde \rho}\in B_2$
into contributions corresponding to a depolarized occupied
space $\tilde O = \sum_{\tilde i}\ket{\tilde i}\bra{\tilde i}$ and
its complement $1-\tilde O$. The depolarized MOs are obtained as
\begin{align}
   \{\ket{\tilde i}\} &= \operatorname{orth}\{P_{12}P_{21}\ket{i}\}
\end{align}
by projecting the accurate MOs $\ket{i}$
from the main basis $B_1$ onto the minimal basis $B_2$ (which cannot
express polarization) and back.\footnote{A projector to space $Y$ maps any point $x$ to its closest point $y\in\mathrm{span}(Y)$, i.e. $P x = \mathrm{argmin}_{y\in Y}\Vert x - y\Vert_2$]}
We can then get the
polarized AOs $\ket{\rho}$ from the free-atom AOs $\ket{\tilde \rho}$
by simply projecting their contributions in $\tilde O$ and $1-\tilde O$ onto their polarized counterparts
$O=\sum_i \ket{i}\bra{i}$ and $1-O$:
\begin{align}
   \ket{\rho} &= \Bigl(O \tilde O + (1-O)(1-\tilde O)\Bigr)P_{12}\ket{\tilde \rho}.\label{eq:ib}
\end{align}

Thus, to construct the polarized AOs it is sufficient
to load a free-atom basis, calculate its overlap with the main basis
and within itself, and perform the numerically trivial projection \eqref{eq:ib}.
Contrary to the related approach of Ref. \onlinecite{Ruedenberg:MoleculeIntrinsicMinimalBasis},
no functional optimization or reference to virtual orbitals is required.
In this article we will also symmetrically orthogonalize the vectors obtained
by \eqref{eq:ib}, to arrive at an orthonormal minimal basis which divides
the one-particle space into atomic contributions;
the latter will be referred to as 
intrinsic atomic orbitals (IAOs).

While the construction makes reference to free-atom AOs
through basis $B_2$, it must be stressed that these are
\emph{not} empirical quantities. Free-atom orbitals can be \emph{calculated} with any high-level
quantum chemistry program.
However, in practice this is not even required
because they are already tabulated as part of several standard basis sets;
here we take the AO functions of cc-pVTZ.\cite{bib:SupplMath}

Since IAOs are directly associated with atoms, they can be used
to define atomic properties like partial charges.
Let us denote the closed-shell SCF density matrix as
$\gamma=2\sum_i \ket{i}\bra{i}$, where $i$ are the occupied MOs.
We can then define
\begin{align}\label{eq:PartialCharge}
   q_A = Z_A - \sum\nolimits_{\rho\in A} \braket{\rho|\gamma|\rho}
\end{align}
as the partial charge on atom $A$, where $Z_A$ is the atom's nuclear charge and
$\rho$ its IAOs.
\tabref{tab:BasisSetDependence} shows that the
partial charges obtained are insensitive to the basis set,
follow trends in electronegativities, and some
defects seen in other methods (e.g., Bader's description of the CN
bond in HCN as ionic) are absent.
Partial charges will be further analyzed below.

\begin{table}
   \begin{tabular}{llcclccc}
   \hline\hline
                     &{\rule{3ex}{0ex}}& \multicolumn{2}{c}{CH$_4$} &{\rule{4ex}{0ex}}& \multicolumn{3}{c}{HCN}
   \\\cline{3-4}\cline{6-8}
        Method/Basis && C & H && H & C & N
   \\\hline
           IAO/def2-SVP$^a$          &  & --0.49 &  +0.12 &    &  +0.21 & --0.01 & --0.20
   \\      IAO/def2-TZVPP$^a$        &  & --0.52 &  +0.13 &    &  +0.22 & --0.01 & --0.21
   \\      IAO/def2-QZVPP$^a$        &  & --0.52 &  +0.13 &    &  +0.22 & --0.01 & --0.21
   \\      IAO/cc-pVTZ$^a$           &  & --0.52 &  +0.13 &    &  +0.22 & --0.01 & --0.21
   \\      IAO/aug-cc-pVTZ$^a$       &  & --0.52 &  +0.13 &    &  +0.22 & --0.01 & --0.21
   \\[1ex] Bader/TZ2P$^b$           &  &  +0.05 & --0.01 &    &  +0.19 &  +0.82 & --1.01
   \\[1ex] Mulliken/DZ$^b$          &  & --0.98 &  +0.25 &    &  +0.34 &  +0.03 & --0.38
   \\      Mulliken/DZP$^b$         &  &  +0.05 & --0.01 &    &  +0.16 &  +0.28 & --0.44
   \\      Mulliken/TZ2P$^b$        &  &  +0.61 & --0.15 &    & --0.02 &  +0.27 & --0.25
   \\\hline\hline
   \end{tabular}
   \caption{(a) Hartree-Fock partial charges
      via \eqeqref{eq:PartialCharge}.
      (b) Kohn-Sham/BP86 partial charges\cite{Guerra:VDD} via
         the Bader and Mulliken methods.
   }
   \label{tab:BasisSetDependence}
\end{table}

IAOs provide access to atomic properties, but
it is often desirable to get a clearer picture of
molecular bonding. We now show that by combining the
IAOs with orbital localization in the spirit of Pipek-Mezey (PM),\cite{pipek:PMlocalization}
one can explicitly construct bond orbitals (IBOs), without any empirical
input, and \emph{entirely within the framework of MO theory}. 
A Slater determinant $\ket{\Phi}$ is invariant to unitary
rotations $\ket{i'} = \sum_i \ket{i} U_{ii'}$ amongst its occupied MOs $\ket{i}$.
We can thus define the IBOs by maximizing
\begin{align}\label{eq:LocFunctional}
   L = \sum_{A}^\mathrm{atoms}\sum_{i'}^\mathrm{occ} [n_A(i')]^4
\end{align}
with respect to $U_{ii'}$. 
Here $n_A(i') = 2\sum_{\rho\in A}\braket{\rho|i'}\braket{i'|\rho}$ is the number of $\ket{i'}$'s electrons
located on the IAOs $\rho$ of atom $A$.
This construction effectively minimizes the number of atoms an orbital
is centered on.
The exponent 4 is preferred over the exponent 2 of PM because it leads to discrete
localizations in aromatic systems; for other systems both exponents lead to effectively identical results.

\begin{figure}
   \centering
   \includegraphics[width=.95\figurerefwidth]{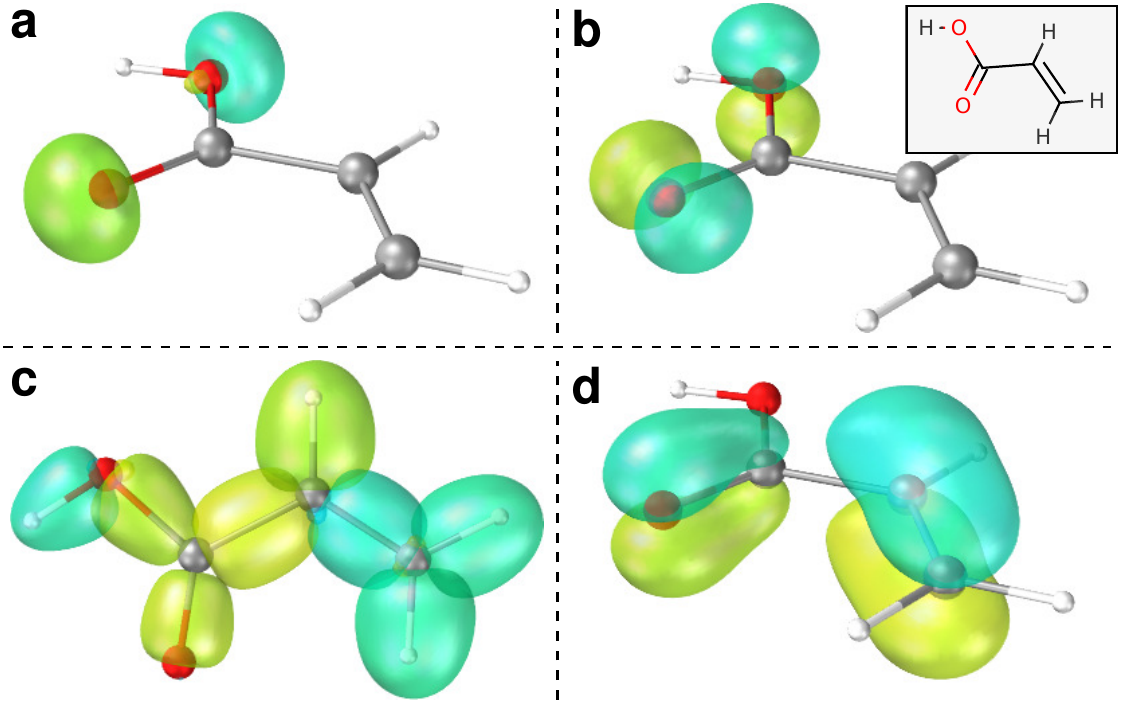}
   \caption{IBOs of acrylic acid: {\bf a} and {\bf b}:
      two $sp$-hybrid and two $p$ lone pairs (1-center orbtials).
      {\bf c} and {\bf d}: eight $\sigma$-bonds and two $\pi$-bonds (2-center orbitals). Core orbitals are not shown.
      }
   \label{fig:AcrylicAcidILO}
\end{figure}

\figref{fig:AcrylicAcidILO} shows IBOs computed for
the acrylic acid molecule.
Here 16 of 19 occupied MOs can be expressed to >$99\%$ by charge
with IAOs on one or two centers, respectively. The three other
MOs are part of a $\pi$-system:
The oxygen $p$ lone pairs (which have about 7\% bonding character),
and the C$=$C $\pi$-bond (which has about 3\% contributions
on the third C atom, and about 1\% on the doubly-bonded O).
In total, we see a direct correspondence of the obtained
IBOs with the classical bonding picture: $\sigma$-bonds, $\pi$-bonds,
and lone pairs are exactly where
expected, and the $\pi$-system is slightly delocalized.
We stress again that these 19 IBOs
are \emph{exactly equivalent} to the occupied MOs
they are generated from: Their anti-symmetrized product is the SCF
wave function, and this is a valid representation of its
electronic structure.
Note that the IBO construction
makes no reference to the molecule's Lewis structure whatsoever;
the classical bonding picture thus arises as an emergent
phenomenon rooted in the molecular electronic structure itself, even if not imposed.

The major improvement of IBOs over PM orbitals is that they are based on IAO
charges instead of the erratic Mulliken charges (cf. CH$_4$ in \tabref{tab:BasisSetDependence}).
As a result, IBOs are always well-defined, while PM orbitals are 
unsuitable for interpretation because they are
unphysically tied to the basis set (they
do not even \emph{have} a basis set limit).
IBOs lift this weakness while retaining and even improving on PM's computational
attractiveness.\cite{bib:SupplMath}

\section{Consistency with empirical facts}
Our hypothesis is that IAOs offer a chemically sound
definition of atoms in a molecule.
But since these are not physically observable,
this claim can only be backed by consistency with empirical laws and facts.\cite{Meister:PrincipalComponentsOfIonicity}
We thus now investigate
whether partial charges derived
from IAOs follow expected trends based on electronegativies,
C 1s core level shifts, and linear free-energy relationships
for resonance substituent effects (Taft's $\sigma_R$).
We then go on to see how IBOs reflect bonding in some non-trivial molecules.

\begin{figure}
   \includegraphics[width=.48\figurerefwidth]{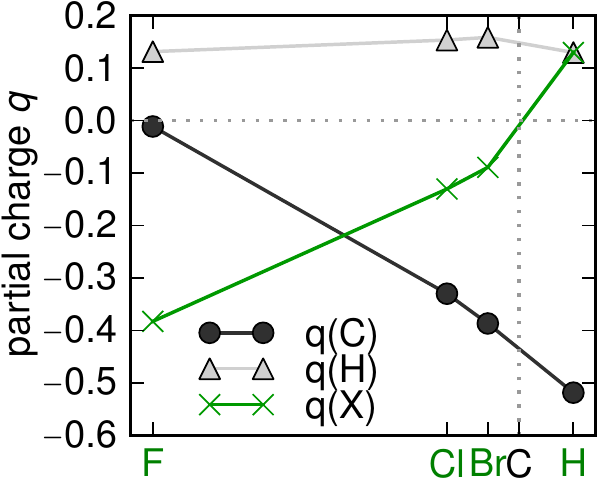}
   \hspace*{0.02\figurerefwidth}
   \includegraphics[width=.48\figurerefwidth]{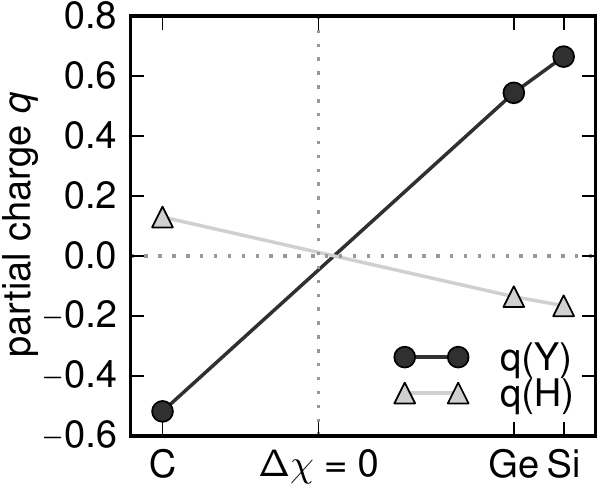}
   \caption{(a): IAO partial charges in CH$_3$X (X=F, Cl, Br, H)
      plotted against $\chi(C)-\chi(X)$. $\Delta \chi=0$ is marked by a dotted line.
      (b): Partial charges in YH$_4$ (Y=C, Si, Ge), the x-axis is $\chi(H)-\chi(Y)$.}
   \label{fig:HalogenMethanes}
\end{figure}

We saw in \tabref{tab:BasisSetDependence} that, unlike Mulliken charges,
IAO charges are insensitive to the employed basis set, 
and unlike Bader charges,
IAO do not erronously describe the CN bond in HCN as ionic.
We now follow Ref.\ \onlinecite{Guerra:VDD} and investigate
IAO charges in relation 
to electronegativity $\chi$ diffences.
We start with the series CH$_3$X (X=F, Cl, Br, H). Due to the
(Allen\cite{Allen:Electronegativities}) electronegativities 
(F: 4.193, Cl: 2.869, Br: 2.685, C: 2.544, H: 2.300),
we expect halogens to have a negative charge, getting smaller in the series,
and hydrogen to have a positive charge.
As shown in \figref{fig:HalogenMethanes}{\bf a},
this is what we find.
In \figref{fig:HalogenMethanes}{\bf b}, we show the series YH$_4$ (Y=C, Si, Ge). We find
charges in close correspondence with $\chi$ (C: 2.544, Si: 1.916, Ge: 1.994),
and the inversion that $\chi(\mathrm{Si})<\chi(\mathrm{Ge})$ is properly reflected.
If we extrapolate the curves in {\bf a} and {\bf b} to $\Delta\chi=0$,
we find in both cases that $q(X)\approx 0$ and $q(Y)\approx 0$, respectively.
That is, if there is no difference in electronegativity, IAO partial charges predict
no bond polarization. This consistency with empirical electronegativies is further
reflected in the almost linear shapes of the curves.
In the series CH$_{4-n}$F$_n$ (n=0$\ldots$4),
we find C partial charges of -0.52, -0.01, 0.44, 0.85, 1.23. The charge increase by $\approx$0.5e$^-$ per fluorine atom
agrees with the understanding of CF bonds in organic chemistry\cite{ohagan:CFbond}
and earlier calculations,\cite{wiberg:cf4}
contrary to the much smaller charges found in Hirshfeld and Voronoi deformation density (VDD) analysis.\cite{Guerra:VDD}

A different test of IAO charges can be performed by comparing to experimental data which are known to be
highly correlated with charge states of specific atoms $A$. A prime
example for this is the C 1s core-level ionization energy shift due
to the molecular environment.
This shift can be estimated\cite{siegbahn:SimplePotentialModel} as
\begin{align}
   \Delta \mathrm{IP}_\mathrm{C\,1s} = k\cdot q_A + \sum_{B\neq A}\frac{q_B}{\norm{\vec R_A-\vec R_B}} + \Delta E_\mathrm{relax},\label{eq:SimplePotentialModel}
\end{align}
where the second term is an estimate for the electrostatic potential of the
other atoms $B$, the last term is a contribution due to core orbital
relaxation, and $k$ is a (hybridization dependent) proportionality constant. 
This model has been employed to calibrate
widely used electronegativity equilibration models,\cite{gasteiger:iterativeequalization}
and has been found to be perfectly correlated with both experimental\cite{guadagnini:CoreEnergiesIrAndCharges}
and theoretical\cite{deOliveira:SimplePotentialModelVsAtomicCharges} mean dipole
derivatives (which for the molecules studied here can be interpreted as charges,\cite{deOliveira:SimplePotentialModelVsAtomicCharges}
but not generally\cite{milani:IrCharges}).
\begin{figure}
   \includegraphics[scale=1.0]{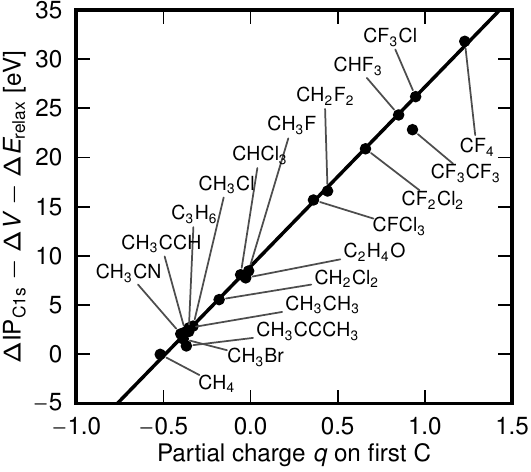}
   \caption{Partial charge on sp$^3$ carbon versus experimental C\,1s
      ionization energy shift corrected for core relaxation effects
      and electrostatic potentials of the other cores (\eqeqref{eq:SimplePotentialModel}), relative
      to methane.}
   \label{fig:C1s}
\end{figure}
In \figref{fig:C1s} we show the results obtained with IAO partial charges
based on Hartree-Fock wave functions for all the sp$^3$ hybridized
molecules as studied in Ref. \onlinecite{guadagnini:CoreEnergiesIrAndCharges}.
IAO charges where used both for the abscissa and $V$, the second term in \eqeqref{eq:SimplePotentialModel}.
$\Delta E_\mathrm{relax}$ was calculated at Hartree-Fock level by a $\Delta$SCF procedure.
We here obtain a linear regression coefficient of $r=0.997$, or $r=0.9995$ if the two
outliers CF$_3$CF$_3$ and CH$_3$C$\equiv$CH$_3$ are excluded. This is the same
level of correlation as obtained with dipole moment derivatives,\cite{guadagnini:CoreEnergiesIrAndCharges}
and much higher than for CHELPS, Bader, or Mulliken charges.\cite{deOliveira:SimplePotentialModelVsAtomicCharges}

\begin{figure}
   \includegraphics[scale=1.0]{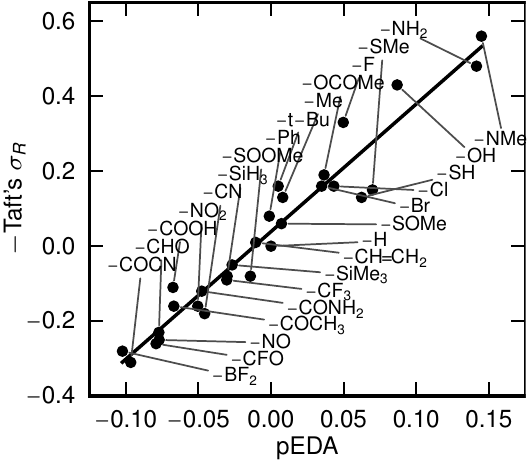}
   \caption{Correlation of pEDA based on IAO/Hartree-Fock charges
      versus $\sigma_R$ substituent constants taken from Ref. \onlinecite{hansch:SubstituentConstantsReview}.}
   \label{fig:pEDAvsSigmaR}
\end{figure}

One advantage of Hilbert-space based partial charges over
real-space partial charges is that they can be split not only
into atomic contributions, but also orbital contributions.
Recently Ozimi\'{n}ski and Dobrowolski\cite{oziminski:EDA} used this freedom
to introduce a set of descriptors for the electronic $\sigma$- and $\pi$-substituent
effects, called sEDA and pEDA, and showed that they are both
internally consistent and highly correlated with empirical
substituent effect parameters like Taft's $\sigma_R$. Concretely,
for a substituted benzene R-C$_6$H$_5$, the pEDA parameter is defined as
the number of $p_z$ electrons on the six carbon atoms of the benzene ring,
relative to the unsubstituted benzene:
\begin{align}
   \mathrm{pEDA} = \sum_{i=1}^6 q_\mathrm{C_i 2 p_z}(\text{C$_6$H$_6$}) - \sum_{i=1}^6 q_\mathrm{C_i 2 p_z}(\text{R-C$_6$H$_5$}),
\end{align}
where Ozimi\'{n}ski defined this quantity based on NAO population
analysis\cite{weinhold:npa} with a specified type of wave function and basis set.
In order to demonstrate the potential of IAO charges in the interpretation of
chemical reactivity, in \figref{fig:pEDAvsSigmaR} we show that the same kind
of correlation with empirical substituent constants is also obtained when
calculating pEDA from IAO charges ($r=0.966$) instead of NAO charges ($r=0.943$\cite{oziminski:EDA}).
Calculating such EDA values is computationally trivial,
which makes them an attractive quantity in the study of unusual substituents
not be contained in common tables, or for testing if the substituents
behave differently for different hosts than benzene.
Indeed, a similar idea to Ozimi\'{n}ski's has been considered previously,\cite{wiberg:SigmaPiMethane}
but was much less practical due to being based on carefully crafted real-space integration
because Hilbert space approaches were considered unreliable.\cite{wiberg:SigmaPiMethane}

A deeper insight into the nature of a molecule's bonding can be obtained
by calculating its bond orbitals.
As previously noted,
IBOs are an \emph{exact} representation of SCF wave functions,
and we have seen in \figref{fig:AcrylicAcidILO}
that they normally reflect the classical bonding concepts
one to one.
However, in many molecules the Lewis structure does not tell the entire
story. Therefore we now probe how IBOs reflect bonding in some well
known, but in different senses non-trivial molecules.
\begin{figure}
   \centering
   \includegraphics[width=.95\figurerefwidth]{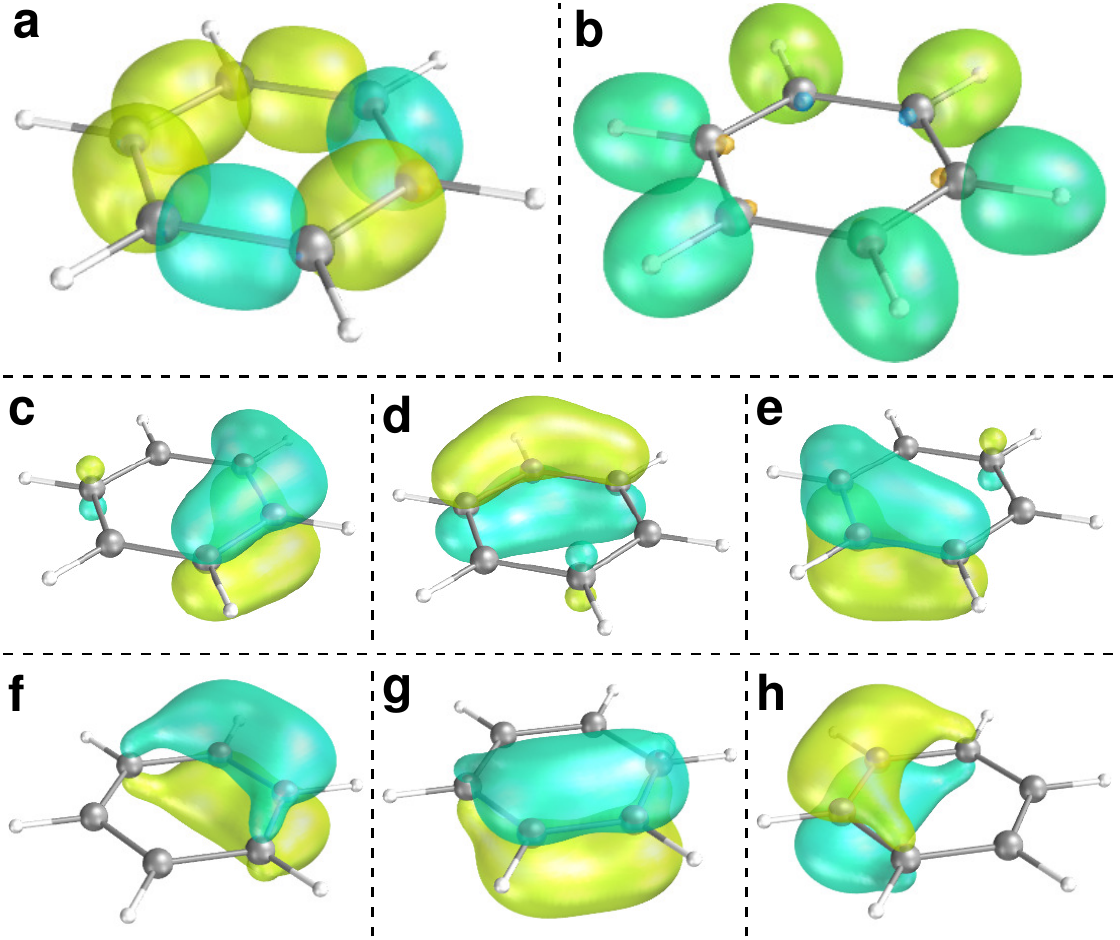}
   \caption{IBOs of benzene: {\bf a} six CC $\sigma$-bonds,
      and {\bf b} six CH $\sigma$-bonds, both localized, and {\bf c-e} one of the two
      equivalent IBO sets representing the delocalized $\pi$-system of three orbitals.
      {\bf f-h}: See text}
   \label{fig:BenzeneIBO}
\end{figure}
\begin{figure}
   \centering
   \includegraphics[width=.95\figurerefwidth]{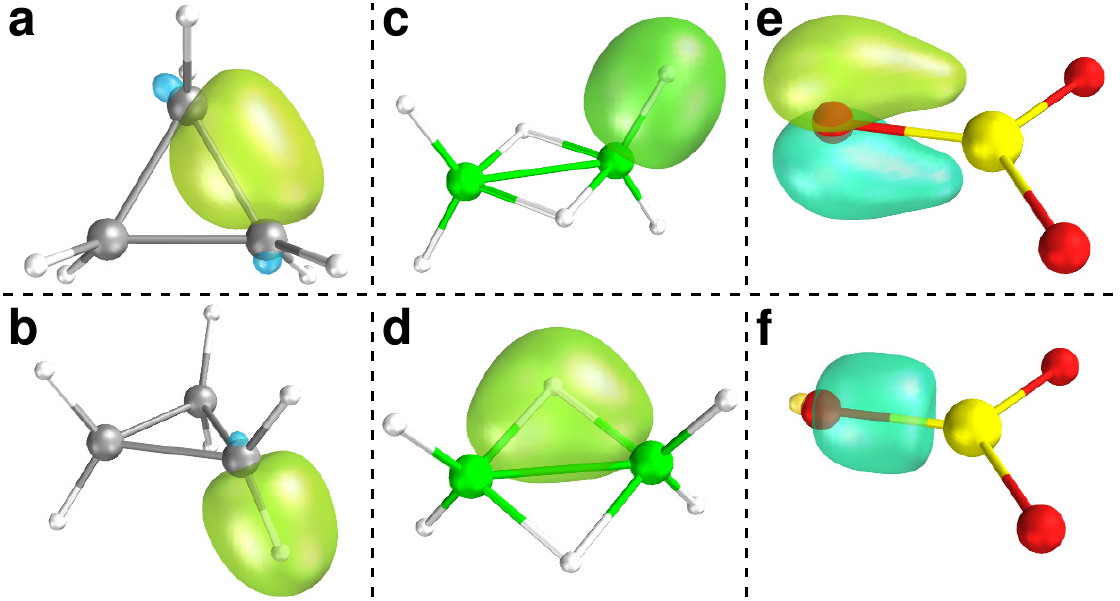}
   \caption{IBO of some molecules with non-Lewis bonding:
     {\bf a} CC banana bond and {\bf b} CH $\sigma$-bond of cyclopropane,
     {\bf c} BH $\sigma$-bond and {\bf d} BHB 2e3c bond of diborane,
     {\bf e} SO $\pi$-bond and {\bf f} SO $\sigma$-bond of sulfur trioxide.}
   \label{fig:OtherIBO}
\end{figure}

Benzene:
A straight application of the IBO construction produces the orbitals {\bf a}--{\bf e} shown in \figref{fig:BenzeneIBO}.
As expected, both the CC and the CH $\sigma$-bonds of the
system are completely localized, and can be be expressed to >99\%
with IAOs on only the two bonded centers. As a prototypical
delocalized system, this does, however, not hold for the $\pi$ system.
Two aspects are important: (i) The three $\pi$ orbitals cannot be
expressed with IAOs on less than four centers each (in {\bf c}--{\bf e} having the weights
1.000, 0.444 (ortho), and 0.111 (para)), and (ii) there are
two different maximal localizations of the functional \eqeqref{eq:LocFunctional},
orbitals {\bf c}--{\bf e} and orbitals {\bf c}--{\bf e} rotated by 60$^\circ$ in real space.
If in \eqeqref{eq:LocFunctional} we had chosen 
to maximize $\sum n_A(i)^2$ instead of $\sum n_A(i)^4$,
there would even be
a continuum of maximal localizations, also including orbitals {\bf f}--{\bf h}
(which look closer to classical $\pi$-bonds), and everything
in between.
In this case the classical resonance structure
reflects the nature of the bonding well.

Cyclopropane:
According to its Lewis structure, cyclopropane is a simple alkane. However, due to 
the massive ring strain one could expect that some
bonding effects are inevitable. Nevertheless, if we calculate the IBOs of this
molecule (\figref{fig:OtherIBO}, {\bf a} and {\bf b}),
we find six CH single bonds and
three CC single bonds, all perfectly localized (to >99\%)
on the two bonded centers, with no delocalization whatsoever.
However, a closer look reveals that while the carbon part of the
CH bond orbitals has about 28\% $s$ character and 72\% $p$ character,
(close to the ideal $sp^3$ hybrid values of $1/4 s + 3/4p$),
the CC bonds only have 18\% $s$ character and 82\% $p$ character.
So although they are localized single bonds,
they must be considered an intermediate between a
regular $sp^3$-hybrid $\sigma$-bond and a $\pi$-bond.
This explains the well-known similarity in reactivity
to alkenes.

Diborane: B$_2$H$_6$ has been
a serious challenge to the classical bonding picture,
with even scientists like Pauling
championing an ethane-like structure
until proven wrong irrefutably.\cite{laszlo:ADiboraneStory}
Its bridged structure was popularized in 1943,\cite{longuet:StructureOfBoronHydrides}
and spawned investigations culminating in
Lipscomb's 1976 Nobel price for his ``studies on the structure of boranes illuminating problems of chemical bonding''.
One could think that this molecule presents a challenge to
a IBO bonding analysis.
However, IBOs are just the most local exact description of a first-principles
wave function, and their construction 
does not make \emph{any} reference to \emph{any} perceived nature of the bonding.
Consequently, for IBO analysis diborane
is not different than other molecules, and it uncovers diborane's
two two-electron three-center bonds (\figref{fig:OtherIBO}{\bf d})
just as its six standard $\sigma$-bonds ({\bf c}), without any problems.

\newenvironment{tightcenter}{%
  \setlength\topsep{3pt}
  \begin{center}
}{%
  \end{center}
}
%
%
Sulfur trioxide: SO$_3$ is one of the simplest ``hypervalent'' molecules,
apparently violating the octet rule.
An IBO analysis finds two oxygen lone pairs, one $\sigma$- and one $\pi$-bond
(\figref{fig:OtherIBO}{\bf e}) per oxygen.
Formally this calls for describing the SO bonds as double bonds.
However, the $\pi$-bonds have only
a small bonding component (83\% on oxygen, 15\% on sulfur),
so it is a matter of taste whether they should be called true $\pi$-bonds or not.
But in any case, they are highly localized (98\% on two centers) and
clearly not resonating, so the resonance structure
\begin{tightcenter}
   \includegraphics[width=2.5in]{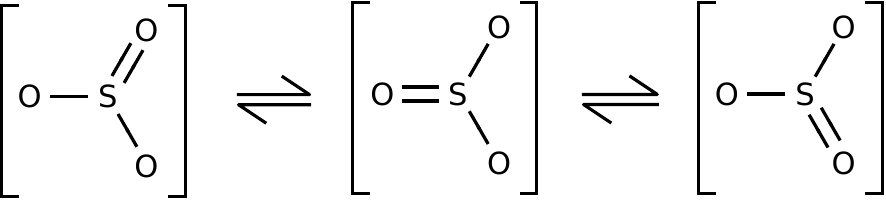}
\end{tightcenter}
---commonly found in textbooks---is at best misleading.\footnote{
Three double bonds are allowed here because they are polarized, and thus contribute less
than one electron per bond to sulfur's valence orbitals. Consequently,
two electrons per valence orbital are never exceeded (the quantum mechanical core of the octet rule) and neither
delocalization nor $d$-orbital participation need to be invoked.}

Bifluoride anion:
We see a similar discrepancy to textbook knowledge
in the description of FHF$^-$. 
This molecule is alternatively cited as the strongest known hydrogen bond,\cite{gilli:UnifiedHydrogenBondTheory} or as an example
for a 3-center 4-electron bond (since an influential paper by Pimentel\cite{pimentel:fhfAnion}).
However, IBO anaylsis reveals that it can be perfectly described by
six F lone pairs and two HF single bonds, all completely localized.
Since the bonds are highly polarized, there is again no violation of the
octet rule: In fact, the H orbital has a population of only 0.6 electrons
total (out of the up to two electrons it theoretically could harbor), and
the nature of bonding in this molecule is not very different than in HF.

\section{Conclusions and Outlook}
The proposed IAOs offers
a simple and transparent way to relate chemical intuition to
quantum chemistry. In particular, the fact that
most simple bonds can be expressed to >99\% with IAOs on only two atoms strongly
indicates that IAOs can be interpreted as \emph{chemical}
valence orbitals in molecules. That properties of individual such orbitals can
then be directly calculated, as shown in \figref{fig:pEDAvsSigmaR},
may turn out to
be a decisive factor in future research on chemical reactivity.
Similarly, IAOs may greatly simplify the construction 
of realistic tight-binding model Hamiltonians
and their use in eludicating complex correlated electronic structure phenomena.\cite{knizia:HubbardDmet,knizia:HydrogenDmet}

The proposed IBOs can help to uncover the nature
of bonding in molecules---due to their unbiased nature also in unusual cases.
However, the IBO construction's simplicity,
ease of implementation, and high runtime efficiency make it
an excellent choice also where localized orbitals are used
for purely technical reasons (e.g., in local electron correlation methods).

We acknowledge funding through ERC Advanced Grant 320723 (ASES).

%
%
%

\end{document}